\documentstyle[12pt]{article}
            \newcommand{\be}{\begin{eqnarray}}
            \newcommand{\ee}{\end{eqnarray}}
            \newcommand{\eel}[1]{\label{#1}\end{eqnarray}}
            \newcommand{\ga}{{\gamma}}
            \newcommand{\al}{{\alpha}}
            \newcommand{\ra}{{\rightarrow}}           
            \newcommand{\ep}{\epsilon}
            
	    \newcommand{\pr}{^{\prime}}
            \newcommand{\halv}{\frac{1}{2}}                                 
            \newcommand{\nn}{\nonumber}
                         
\begin{document}
            \begin{titlepage}
            \newpage
\noindent
                CWRU-96-P8\\
                May 1996\\

            \vspace*{10 mm}
            \begin{center}{\large\bf HIGH ENERGY PROTON ACCELERATION 
IN DISSOLVING PLASMA CLOTS}\end{center}
           \vspace*{10 mm}
        \begin{center}{ \bf V. DAVIDOVSKY and  
\bf SH. SHVARTSMAN 
\footnote{e-mail: sms25@po.cwru.edu}}
          \vspace*{10 mm} \\
{\sl  Physics Department  \\
Case Western Reserve University\\ 
          Cleveland, OH 44106, USA }
\end{center}
\begin{abstract}
We consider the acceleration of particles (protons) in a plasma 
clot by an induced field which appears  due to the dissolving of 
the clot. The mechanism of this acceleration is based on a  
model where the clot consists of  double
flat moving current layers. It is
assumed that the currents flow in opposite directions,
while the clot moves in a perpendicular one. It is shown that
protons can reach the highest
observed energy (of the order of $10^{11}- 10^{12}$GeV), providing 
that the size of the clot is large enough.
\end{abstract}
\end{titlepage}

\setcounter{equation}{0}

	There is a well known problem in high energy cosmic rays physics
of how one can explain the existence in the universe of protons with the
energy in the range of $10^{11}-10^{12}$GeV. Unfortunately all existing 
models do not successfully describe this phenomena. 

	In this paper we are going to discuss a mechanism that might
explain the existence of particles which have  such high energies. 
This mechanism assumes that somewhere in the universe there exist 
very large and sufficiently dense plasma clots. 
Such objects recently were discovered in space, indeed these are 
the SS433 objects \cite{Stephenson} and the jets from the active 
galactic nuclei \cite{Pearson}. We also assume that inside the clots there
exists an internal current as it will be described below.

	Plasma clots usually posses collective electromagnetic field. 
This field may be  produced  by either a collective electric charge  
or a current inside  the clot. The collision of the clot's particles 
  with the medium,
as well as the internal recombination in the cooling clot, are
responsible for dissolving of the moving clot. 
Therefore, the collective 
electromagnetic field decreases, and the energy and the momentum 
of the vanishing field can be distributed among the rest of the clot 
particles. This means that the energy of the collective field can 
be transferred to 
a comparatively small number of particles, and 
those particles can reach high energy. 
This effect might explain the origin of cosmic 
ray protons with the highest observed energies.
Recently, some events were discovered where the energies are 
larger than $10^{11}$ GeV \cite{Bird,Bird1,Brooke,Efimov}. 

	As the first step toward the justification of the above 
proposed acceleration mechanism ,  we shall show that the energy 
spectrum of  accelerating particles is close to the observed one.  
For this purpose it is natural
to assume that the number of particles, say $N_p$, in the clot decreases 
exponentially in time
\be
N_{p}=N_{0\;p} \exp(-\al t)
\eel{1.1}
where  $N_{0\;p}$ 
is the initial number of particles in the clot and $\al$ is a slowly 
varying function of time. Then, during the time interval 
$dt$ the loss in the clot energy, $dE$, can be presented as
\be
dE =\ep dN_{p} 
\eel{1.2}
where $\ep$ is the sum of the kinetic energy of a particle and a 
portion of the collective field energy per one particle at a given 
instant of time. A part of energy $dE$ is transferred to the rest 
of the clot particles, such that each of them receives the energy 
portion $d\ep$
\be
d\ep =-k \frac{dE}{N_{p}}=k\al\ep dt
\eel{1.3}
where k is a proper fraction.  This gives
\be
\ep =\ep_0 \exp(\al k t)
\eel{1.4}
Then a simple relation holds
\be
\frac{N_{p}}{N_{0\;p}}=\left(\frac{\ep}{\ep_0}\right)^{-\frac{1}{k}}
\eel{1.5}
This spectrum is close to the observed one \cite{Hillas} if the parameter k 
is of the order of one-half to two-thirds. This shows that a $k$-portion 
($k\simeq 0.5$)  of the energy of the clot's collective field per one 
particle can be transferred to the rest of the  particles inside the clot. 

	To estimate the maximum of the energy the particles (protons)
can gain, let us consider a simple model as follows. We assume that 
the collective electromagnetic field is produced by a current inside 
the clot. 
This current consists of the two double 
flat current layers flowing in opposite directions, say $x$-direction. 
The current inside the single layer is composed by both the electrons 
that have the velocity $v^{(e)}_x$ and protons that have the 
velocity $v^{(p)}_x$. 
We also suppose  that both layers as whole system is moving in the 
perpendicular direction, say the $z$-one, with the velocity $v_z$ 
relative to the medium. 

	Due to the above proposed model the 
current density in the reference frame $K\pr$ where the current 
layers are at rest can be expressed in the 
following form
\be
j\pr _x =\left \{\begin{array}{l}
j\pr,\;\;   {\rm if}\;\;0<z\pr <L\pr \\
-j\pr ,\;\; {\rm if}\;\; -L\pr <z\pr <0 \\
0,\;\; {\rm if}\;\; |z\pr |> L\pr
\end{array}\right.  ,\;\;\;\;
j\pr _y =j\pr _z =0
\eel{1.6}
where $L\pr$ is the effective width of the layer.

The corresponding vector-potential $\vec{A}\pr$ and magnetic 
field $\vec{B}\pr$ are (we present only the none zero components )
\be
A^{'}_x & =&\left\{\begin{array}{l}
\mu_{0}j^{'}z\pr  (L^{'} -z\pr  /2),\;\;{\rm if}\;\;0<z\pr  <L\pr  \\
\mu_{0}j^{'}z\pr  (L\pr  +z\pr  /2),\;\;{\rm if}\;\;-L\pr  <z\pr  <0 \\
\mu_{0}j^{'}L^{'2}/2,\;\;{\rm if}\;\;z\pr >L\pr  \\
-\mu_{0}j^{'}L^{' 2}/2,\;\;{\rm if}\;\;z\pr <-L^{'}
\end{array}\right. 
\label{1.7}\\
B\pr _y & =&\left\{\begin{array}{l}
\mu_0 j\pr (L\pr -z\pr ),\;\;{\rm if}\;\;0<z\pr <L\pr \\
\mu_0 j\pr (L\pr +z\pr),\;\;{\rm if}\;\;-L\pr <z\pr <0 \\
0,\;\;{\rm if}\;\;|z\pr |>L\pr
\end{array}\right.
\eel{1.8}

	Due to the collisions with the medium particles part of the 
electrons and protons  inside the current layers live the clot. Because the 
plasma clot remains quasineutrality both number of particles 
that leave the clot
are the same. Therefore  current density $j\pr _x$ decreases.  
In the reference frame
$K\pr$ medium particles have 
the velocity $v_z$ directed opposite to the z-axis.
This velocity $v_z$ is the one of the 
current layer in the laboratory reference frame $K$ where the medium is 
at rest. In the reference frame $K\pr$ the concentration of the
clot electrons 
$n\pr$ inside the layer can be written in the following form
\be 
n\pr =n\pr _{0} \exp(-\sigma N\pr v_{z}t\pr)
\eel{1.9}
where $n\pr _0$ is the initial concentration of the clot electrons in the 
layer, $N\pr $ is the concentration of the particles  in the medium, and 
$\sigma $ is the scattering cross section of an electron inside the 
clot by the  medium particles. 
Here we neglect the thermal  motion of the electrons inside the clot.

	Let us now pass to the laboratory reference frame K. 
The component $A\pr _x$ of the vector potential (\ref{1.7}) does 
not transform because it is the transverse component of a 4-vector 
(as well as the current density $j $), although the variables it 
depends upon, do. Then this vector potential in the reference 
frame K has the form
\be
A^{\pm}_x &=&
\frac{\mu_0}{2}j(z-v_z t) [2L\mp (z-v_z t)]\ga^{3}_z,
\label{1.10}\\
j &=&en\pr _0 v_x \exp\{-\sigma Nv_z (t-\frac{v_z}{c^2}z)\ga^{2}_z \},
\label{1.11}\\
\ga_z&=& (1-\frac{v^{2}_z}{c^2})^{-\halv}
\eel{1.12}
where the sign $+$ corresponds to the region 
$0< z-v_z t <L$, while the sign $-$ corresponds to the 
region $-L <z-v_z t <0$. The corresponding electric 
and magnetic fields are (non zero components only)
\begin{eqnarray}
B^{\pm}_{y}&=&B^{\pm}_{y\,a}+B^{\pm}_{y\,c}
\label{1.13}\\
B^{\pm}_{y\,c}&=&\mu_0 j[L\mp(z-v_zt)]
\ga^{3}_z,\;\;\;
B^{\pm}_{y\,a}=j\mu_{0}\frac{v^{2}_{z}}{2c^2}\sigma N (z-v_z t)
[2L\pm (z-v_z t)]\ga^{5}_z ,\nn\\
E^{\pm}_{x}&=&E^{\pm}_{x\,a}+E^{\pm}_{x\,c},
\label{1.14}\\
E^{\pm}_{x\,c}&=&v_zB^{\pm}_{y\,c},
\;\;E^{\pm}_{x\,a}=\frac{c^2}{v_z}B^{\pm}_{y\,a}\nn
\end{eqnarray}
The fields $E^{\pm}_{x\,c}$ and $B^{\pm}_{y\,c}$ are 
the usual fields produced by the moving current, while the 
fields $E^{\pm}_{x\,a}$ and $B^{\pm}_{y\,a}$ are the ones 
which appear due to the dissolving of the clot. 
The acceleration effect we consider here (acceleration of the 
clot particles) is due to  the fields $E^{\pm}_{x\,a}$ and 
$B^{\pm}_{y\,a}$.

	We estimate the maximum of the energy the particles can 
gain due to the described above acceleration. Notice, that the potential 
(\ref{1.10}) does not depend on the transverse coordinate $x$. 
Therefore there exists an integral of motion
\be
p_x +eA_x =p_{0 x}+eA_{0 x}={\rm const.}
\eel{1.15}
where the index $0$ labels the initial condition. As 
$j\ra 0$, the potential $A_{x}$ vanishes as well. 
This allows one to find the maximum of the transverse momentum
\be
(p_{x})_{{\rm max}}= p_{0 x}+eA_{0 x}
\eel{1.16}
	Let us consider now the longitudinal motion of the particles. 
The magnetic field $B^{\pm}_{y a}$ produces the Lorentz force
\be
F_z =ev_{x}B^{\pm}_{y\, a}
\eel{1.17}
that acts on the electrons as well as on the protons in the 
positive z-direction in both regions $z-v_{z}t >0$ 
and $z-v_{z}t <0$.
It should be noticed that the field $B^{\pm}_{y\;a}$ is responsible
for the acceleration of the clot as the whole system in $z$-direction.
Some evidence of a such acceleration were mentioned in \cite{Jaffe}.
  
	Notice that the longitudinal 
force which is acting on the electrons is greater than the one 
acting on the protons. This is because the transverse velocity $v_x$
of the electrons is greater than the one for the protons. Due to the 
quasineutrality of the clot, there appears a longitudinal electric field 
$E_z$ that decelerates the electrons and accelerates the protons.

	The longitudinal equations of motion for both electrons (e) and
protons (p) are
\begin{eqnarray}
\frac{dp^{e}_{z}}{dt}&=&-eE^{\pm}_z +ev^{e}_{x}B^{\pm}_{y\, a},\\
\label{1.18}
\frac{dp^{p}_{z}}{dt}&=&eE^{\pm}_z +ev^{p}_{x}B^{\pm}_{y\, a}\nn
\end{eqnarray}
As a result, one obtains
\be
\frac{d(p^{e}_{z}+p^{p}_{z})}{dt}=e(v^{e}_{x}+v^{p}_{x})
B^{\pm}_{y\, a}
\eel{1.19}
The longitudinal force in eq.(\ref{1.19}) is 
$v_{z}(v^{e}_{x}+v^{p}_{x})/2c^2$
times less than the force $2eE^{\pm}_z$ which is acting on the $e^{-}p$ 
pair in the current direction. Therefore the acceleration in 
the $x$-direction is at least not less than the one in the $z$-direction.
This allows one to estimate the maximum of the proton energy
\be
\ep_{{\rm max}}\approx (p_x)_{{\rm max}}c=c(p_{x 0}+eA_{x 0})
\eel{1.20}
If the increase in the energy is much larger than the 
initial energy, then one finds
\be
\ep_{{\rm max}}\approx ecA_{x 0}\approx \frac{\mu_0}{2}ecj_{0}
(z_{0}-v_{z 0})(2L_0 \mp (z_{0}-v_{z 0}t))\ga^{3}_{z 0}
\eel{1.21}
where $j_0$ is the initial current density. 

One can accept an estimation
\be
(z_0 -v_{z 0}t)(2L_{0}\mp(z_0 -v_{z 0}t))\ga^{2}_{z0}\approx L^{'2}_{0}
\eel{1.22}
This leads to an estimation for the maximum energy
\be
\ep_{{\rm max}}\approx \frac{\mu_{0}}{2}ecj_{0}L^{' 2}_{0}\ga_{z 0}
\eel{1.23}

	Thus we can write an estimation for the maximum gained energy 
\be
\ep_{{\rm max}}\approx 2j_{0}L^{' 2}_{0}\ga_{z\;0}10^{-7}({\rm Gev})
\eel{1.26}
The relatively new evidences \cite{Zwitter} and models \cite{Gomez} 
show that the values
\begin{eqnarray*}
L\pr _{0}\approx (0.01 - 0.1){\rm pc}
\end{eqnarray*}
do not appear to be unrealistic. Accepting this estimation we can find
the current density $j_0$ which is necessary for the protons to gain
the maximum energy of the order of $10^{11}$Gev. This gives
\be
j_0 \approx 5(10^{-11} - 10^{-12}){\rm A/m^{2}}
\eel{1a}
Estimate now the initial magnetic field which corresponds to 
the current
density (\ref{1a}). Assuming that the relativistic factor $\ga_{z\;0}$
is of the order of one we get from (\ref{1.13}) the following
relation
\be
B_0\approx \mu_{0}j_{0}L_{0}
\eel{1b}
This yields
\be
B_{0}L_{0}\approx \frac{\ep_{{\rm max}}}{ec}
\eel{1c}
For $\ep_{{\rm max}} \approx 10^{11}$Gev and $L_{0}\approx 0.1$pc the
necessary value of the magnetic field is about 3Gs.

	These estimations do not depend on the model. 
In order to prove this, we consider two different models\\
1) {\bf Thin long coil}. The vector potential which describes 
the current at the clot particles location has the form
\be
A_{\psi}=\frac{\mu_{0}}{2}jar
\eel{1.24}
where $r$ is the radius of the coil, and $a$ is its thickness. \\
2) {\bf  Thin round current loop}. The vector potential 
describing the current  at the clot particles location has the form
\be
A_{\psi}=\frac{\mu_{0}}{2}ja^{2}\ln (8R/a)
\eel{1.25}
where $a$ is the radius of current string, and $R$ the radius 
of the loop.

It can be shown that in these models the result (\ref{1.23}) 
for the maximum energy has a similar form if one makes the 
following substitutions
\begin{eqnarray*}
L^{' 2}_{0}\ra ar\,   {\rm for\;the\; model}\;\;(\ref{1.24})\nn\\
L^{' 2}_{0}\ra a^{2}\ln (8R/a)\;{\rm for\;the\; model}\;\;(\ref{1.25})
\end{eqnarray*}

	Similar effects appear when the collective field inside the clot
is produced 
by an effective electric charge in the clot instead of the current.
	
{\bf ACKNOWLEDGMENT}

We would like to thank Boris V. Vayner, Lawrence Krauss and the members of 
particles/astrophysics seminar at CWRU for critical  and useful comments. 
We also acknowledge the industrial problem solving
group at CWRU for support of this work. 
We thank Hiro Fujita for careful reading of the manuscript and
critical comments.

\end{document}